\documentclass[aip,groupedaddress,reprint]{revtex4-1}

\usepackage{graphicx}
\usepackage{dcolumn}
\usepackage{bm}

\usepackage[utf8]{inputenc}
\usepackage[T1]{fontenc}
\usepackage{mathptmx}
\usepackage{amsmath}
\usepackage{amssymb}
\usepackage{xcolor}
\usepackage{soul}
\usepackage{gensymb}

\begin{document}

\preprint{AIP/123-QED}

\title[Structural anomalies in brain networks induce dynamical pacemaker effects]{Structural anomalies in brain networks induce dynamical pacemaker effects}

\author{I. Koulierakis}
 \email{koulyia@gmail.com}
\affiliation{ 
Institute of Nanoscience and Nanotechnology, National Center for Scientific Research ``Demokritos'', 15341 Athens, Greece \\
and School of Electrical and Computer Engineering, National Technical University of Athens, 15780 Athens, Greece}%
\author{D. A. Verganelakis}%
 \email{dimitris.verganelakis@gmail.com.}
\affiliation{ 
Nuclear Medicine Unit, Oncology Clinic ``Marianna V. Vardinoyiannis - ELPIDA'', Childrens' Hospital ``A. Sofia'',  11527 Athens, Greece%
}%
\author{I. Omelchenko}
\email{omelchenko@itp.tu-berlin.de.}

\affiliation{%
Institut f\"{u}r Theoretische Physik, Technische Universit\"{a}t Berlin, Hardenbergstrasse 36, 10623 Berlin, Germany}%
\author{A. Zakharova}
\email{anna.zakharova@tu-berlin.de.} 
\affiliation{%
Institut f\"{u}r Theoretische Physik, Technische Universit\"{a}t Berlin, Hardenbergstrasse 36, 10623 Berlin, Germany}%
\author{E. Sch\"{o}ll}
 \email{schoell@physik.tu-berlin.de.}
\affiliation{%
Institut f\"{u}r Theoretische Physik, Technische Universit\"{a}t Berlin, Hardenbergstrasse 36, 10623 Berlin, Germany}%
\author{A. Provata}%
 \email{a.provata@inn.demokritos.gr.}
\affiliation{ 
Institute of Nanoscience and Nanotechnology, National Center for Scientific Research ``Demokritos'', 15341 Athens, Greece}%

\date{\today}

\begin{abstract}
Dynamical effects on healthy brains 
and brains affected by tumor are investigated via numerical simulations. 
The brains are modeled as multilayer networks consisting of neuronal oscillators, 
whose connectivities are extracted from Magnetic Resonance Imaging (MRI) data. 
The numerical results demonstrate that the healthy brain presents chimera-like states where regions
with high white matter concentrations in the direction connecting the two hemispheres
act as the coherent domain, while the rest
of the brain presents incoherent oscillations. To the contrary, in brains with destructed 
structure 
traveling waves are produced initiated at the region where the tumor is located. These areas act
as the pacemaker of the waves sweeping across the brain. The numerical simulations are performed
using two neuronal models: a) the FitzHugh-Nagumo model and b) the Leaky Integrate-and-Fire model. Both
models give consistent results regarding the chimera-like oscillations in healthy brains
and the pacemaker effect in the tumorous brains.
These results are considered as a starting point
for further investigation in the detection of tumors with small sizes
before  becoming discernible
on the MRI recordings as well as in tumor development and evolution.
\end{abstract}

\maketitle
\everymath{\displaystyle}
\begin{quotation}

The detection of human brain anomalous structures (lesions) relies mainly on data from  
Magnetic Resonance Imaging (MRI) or Computed Tomography (CT) scans \cite{gao:2013}.
These techniques are able to discern abnormalities of sizes as low as a few
millimeters. For all medical purposes
it is important to develop effective tools for 
the early diagnosis and detection of tumors of smaller sizes
and  in recent years
many studies have been devoted to exploring early warning indicators 
of abnormal tissue development \cite{pratt:2018,galldiks:2019}.
In this study, we propose an alternative detection method based
on numerical integration of dynamical systems on a network extracted 
from the structure of the white matter as displayed by MRI data. 
The abnormal tissue development generates traveling waves 
whose origin is located
on the lesion which acts as the ``pacemaker''. This is an interesting finding and needs to
be further explored with a dual goal: a)   the determination of the anomalous tissue location
at the center of the pacemaker region and b) the possibility to be useful in relation to 
lesion/tumor detection
at the early stages, before being visible by human eye in the MRI images.
\end{quotation}


\section{Introduction}
\label{Introduction}

Synchronization phenomena are omnipotent in systems consisting of many coupled
oscillatory units and are of great importance in the exchange of information in networks of neurons
which perform nonlinear, spiking oscillations \cite{boccaletti:2018,velazquez:2009,izhikevich:2007}. In fact, neurons
operate as potential integrators with a cutoff, after which they exchange information in the form of
 electrical or chemical signals \cite{gerstner:2002,kandel:2013}. In collective neuron dynamics interesting
synchronization phenomena arise, such as local phase or amplitude synchronization, incoherent and chaotic
oscillations and chimera states. Many of these phenomena are associated with healthy brain functionality 
(intermittent oscillations and chimera states), while others appear as a result of brain malfunctions 
such as epilepsy, schizophrenia 
and neurodegenerative brain disorders \cite{andrzejak:2016,mormann:2000,olmi:2019,chouzouris:2018}. 
In the present study, 
numerical simulations are used 
to investigate and compare synchronization phenomena in healthy brains and brains
affected by tumors. Our approach consists of using structural MRI data
to construct
 the underlying network substrate where the neuronal oscillators operate and interact.
``Chimera-like'' states and traveling waves are some of the most prominent synchronization properties
that are reported in this study in relation to healthy and tumorous brains, respectively.

\par Chimera states are stable states in systems of coupled nonlinear
oscillators which are characterized by coexistence of synchronous and asynchronous
domains. Depending on the system parameters, multichimera states may also be developed
which include many alternating  coherent and incoherent domains. Single chimeras
were first observed by Kuramoto and Battogtokh in 2002 \cite{kuramoto:2002,kuramoto:2002a},
while the term ``chimera'' was introduced two years later, by Abrams and Strogatz 
\cite{abrams:2004}.
The coexistence of synchronous and asynchronous domains is not a trivial effect in view of the fact that
it is often observed in systems of identical oscillators, identically linked and is, therefore,
considered as a spontaneous spatial symmetry breaking phenomenon. Chimera states coexist with
the fully synchronous state and, for finite  system sizes, the chimera may transit into the 
synchronous state.

\par Besides earlier studies which focused on coupled phase
oscillators, the well known Kuramoto model \cite{kuramoto:2002},
later works have reported chimera states in coupled 
oscillators with different dynamics, such as
the Hodgkin-Huxley (HH) model,
the FitzHugh-Nagumo (FHN), the Hindmarsh-Rose,  the Van der Pol,  the Stuart-Landau and the
Leaky Integrate-and-Fire (LIF) oscillator networks 
\cite{sakaguchi:2006,omelchenko:2013,omelchenko:2015,schmidt:2017,hizanidis:2014,ulonska:2016,olmi:2015,tsigkri:2017,tsigkri:2018,santos:2018,gjurchinovski:2017,zakharova:2017,shepelev:2019}.
Recent advances in the field of chimera states and, more generally, on network synchronization are
summarized in references 
\cite{panaggio:2015,schoell:2016,zheng:2016,majhi:2018,oomelchenko:2018,zakharova:2020}. 

\par Experimentally, chimera states have been realized in diverse systems consisting of oscillatory
units, such as in  optical systems \cite{hagerstrom:2012},
catalytic systems \cite{tinsley:2012,wickramasinghe:2013,schmidt:2014}
coupled metronomes \cite{martens:2013,kapitaniak:2014}, 
electronic circuits \cite{gambuzza:2014} and biomedicine
\cite{cherry:2008,mormann:2000,mormann:2003a,santos:2017}.
Future applications are extensively discussed in the field of 
metamaterials \cite{lazarides:2015,hizanidis:2016} and in
biomedical applications \cite{andrzejak:2016,ruzzene:2019,bansal:2019,hizanidis:2015,tsigkri:2016,ulonska:2016}.
In nature, chimera states have been associated with the uni-hemispheric sleep in dolphins and birds 
 as well as the synchronous - asynchronous firing in fireflies
\cite{rattenborg:2000,rattenborg:2006,ramlow:2019}.

\par In relation to biomedical applications, synchronization phenomena are important in brain dynamics in the
exchange of electrical and chemical signals between brain neurons \cite{bera:2019}. That is why in the original studies
of synchronization Kuramoto used the phase oscillator, a prototype model of neuronal firing
 \cite{kuramoto:2002}.
In realistic brain simulations, where different types of neurons are involved with different connectivities,
the classical chimera states cannot be seen. However, even in these cases of inhomogeneous neuronal populations
specific synchronization patterns which include coherent and incoherent domains can be observed and these are
called ``chimera-like states'' \cite{santos:2017} to keep the connection with the classical chimeras where all
oscillators are identical and identically linked. In the present study,  an intermediate path is used: 
coupled identical oscillators (FHN or LIF units) are employed with  
nonidentical connectivities, extracted from the 
intensity of the MRI images of the healthy and tumorous brains. 

\par This work is organized as follows. The MRI 
data of the control subjects and patients, used later on to construct the connectivity matrices, 
is presented in the next section. Section~\ref{sec:models} introduces
the mathematical framework of the coupled FHN and LIF
neuronal network models, while in Sec.~\ref{sec:connectivity}
the realization of the connectivity schemes is presented.
In Sec.~\ref{sec:results}, by means of numerical simulations the emergence
of chimera-like states is discussed in healthy brain 
(Secs.~\ref{results:healthy-FHN},~C) and in
tumorous brain (Secs.~\ref{results:anomalous-FHN},~D),
using two different models:
coupled FHN (Secs. ~\ref{results:healthy-FHN},~B) and LIF (Secs. ~\ref{results:healthy-LIF},~D) oscillator networks. 
The spectral properties of specific oscillators located in the healthy and tumorous areas are discussed
and compared in Sec.~\ref{sec:spectral}.
 The final section, \ref{sec:conclusions}, recapitulates our general conclusions.

\section{The Data}
\label{sec:data}

\par Brain MRI scanning is a noninvasive imaging 
diagnostic method which combines 2D images to create a 3D picture 
of the brain and has been the basic tool for physicians in detecting abnormalities 
in  brain structures  
\cite{bihan:2001,basser:1994,basser:1994b,sugahara:1999,basser:2000,mori:2002}. 

It is a method used since the 1970's and has a wide range of contrast mechanisms
allowing to detect structural and functional brain attributes. One of the most 
popular mechanisms is based on the
relaxation properties of the magnetic moments of water molecules when they get exposed 
to an external magnetic field and oscillating radio waves. Diffusion-weighted magnetic 
resonance imaging (DWI) \cite{sugahara:1999}
 generates maps of the diffusion processes (Brownian motion) 
of water molecules in biological tissues driven by thermal agitation using three 
gradient directions, (x,y,z), where the signal intensity of each voxel represents 
the optimal of local water diffusion rate. Biological tissues are structurally rich  
environments that consist of macromolecules, membranes and fibers, where the actual 
diffusion process of water molecules is affected by the architecture of those components,
 depicting macroscopically their architecture. 
\par  The Data of the present study has been obtained via
the Diffusion Tensor Imaging (DTI) technique \cite{basser:2000,sipkins:1998,mori:2007}, 
which is a direct extension of DWI.
It is a technique that maps white matter structure 
in brain using six or more gradient directions allowing to calculate the diffusion tensor. 
The signal intensity of each voxel encapsulates both the local diffusion rate and 
the major local diffusion direction demonstrating the 3-dimensional shape of 
the structure. Fibers’ directions are denoted by the tensor’s main eigenvector. 
Color-coded main eigenvectors produce maps of bundles of axonal neurons with respect to position and direction.
In white matter, 
water molecules diffuse more freely along the bundles of neuronal axons in 
the white matter of brain capturing their 3D orientation, structural integrity and concentration. This preferentially oriented diffusion is called anisotropic diffusion. 
Diffusion anisotropy measures, such as fractional anisotropy (FA), can be derived from the diffusion tensor.
FA takes values between 0 
and 1, and  reflects fiber density, and axonal diameters in white matter. 
It is used here to 
represent the local axons density, allowing to transfer electrical signals (in the
form of potential variations) between the different parts of the brain.

\par For our study 3D MRI data from 4 subjects were considered:
two healthy subjects, denoted as H(1) and H(2) and two patients with brain tumor,
denoted as P(1) and P(2).
For each subject the data consist of a set of RGB (red-green-blue) images
which, using the water diffusion in 3D space, depict the density
of the neuron axons locally in the brain. 
Among other data treatments during the post-processing, the scalp  and skin
of the subjects was computationally removed.
The number of RGB slices $n_s$ ranges between $40 \le n_s \le 44$ for each subject; 
the slices are taken equidistantly along the
z-axis of the brain (superior-to-inferior direction). Each slice
consists of 256 $\times$ 256  cells.
All 2D slices are stacked algorithmically to reconstruct the 
3D brain structure, which ultimately consists of 
256 $\times$ 256 $\times \> n_s$ voxels.
The color scale (minimum=0, maximum=1, in arbitrary units) denotes
the density of neuron axons within each voxel.

\par  All images are obtained from a General Electric 1.5 Tesla Signa
HDxt MRI Scanner.
 The coil that is used to transmit the
radiofrequency pulses and detect the MRI signal is an
eight-element head coil. The parameters of the Diffusion Tensor Imaging (DTI) single-shot spin-echo Echo Planar Imaging (EPI) pulse sequence
are: flip angle 90\degree, Echo Time (TE): 85 ms,
Repetition Time (TR): 10,700 ms, slice thickness: 3 mm,
spacing between slices: 0 mm, Field Of View (FOV):
26 cm, matrix: 256 $\times$ 256, and No. of Excitations (NEX):
1. The voxel resolution was 3 mm $\times$ 3mm $\times$ 3 mm. The b
value used is 1000 s/mm$^2$ and the diffusion gradients
are applied along 30 non-collinear directions. The
scan time for each subject is approximately 5 min. The
number of directions is a compromise between resolution
and acquisition time; the time required for a larger number
of directions would necessitate longer scan time, with
higher probability for artefacts in the data due to erratic motion
of the scanned subjects. The data has been previously published in Refs. 
\cite{katsaloulis:2009,katsaloulis:2012,katsaloulis:2012b,provata:2012}.

\par
\begin{figure}
\includegraphics[width=0.49\textwidth,angle=0.0]{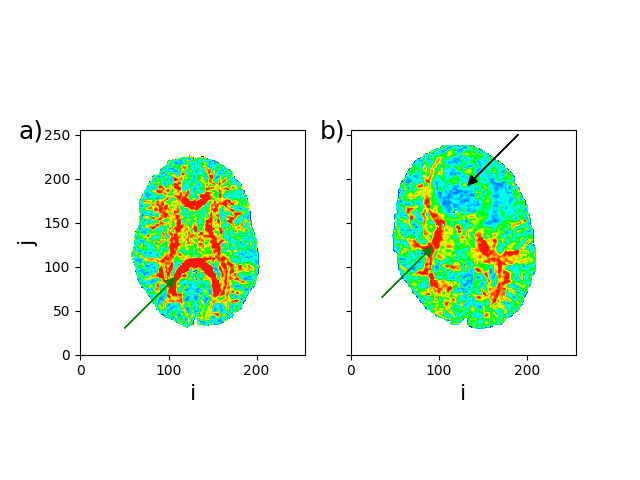}
\caption{\label{fig:01} MRI slices of a) a healthy brain (slice 19 out of 44,  control H(1))
and b) a tumorous brain (slice 19 out of 43, patient P(1)).
The color grading represents the fractional anisotropy values, i.e. local density of neuron axons.
The green arrows point to the corpus callosum areas and the black arrow in (b) to the 
tumorous area.}
\end{figure}

 Figure \ref{fig:01} depicts representative MRI slices in the middle of the vertical
axis through the brain. The left panel a) depicts slice 19 (out of 44 slices) of the
 healthy control subject H(1), and the right panel 
b) depicts slice 19 (out of 43 slices) of patient P(1) with brain tumor.
The color intensity depicts the local axons density and direction
(Red: left-right,  
Green: anterior-posterior,
Blue: superior-inferior).

\par In the healthy subject we note the presence of four red ``ribbons'' crossing the brain structure,
symmetrically. These structures constitute the corpus callosum regions which
are characterized by high white matter concentration in the transverse plane
(see green arrow in Fig.~\ref{fig:01}a).
These regions will be referred to as cc ribbons or cc areas. We call ``Set $A(H)$'' 
the area covered by the cc ribbons in the healthy control and 
 ``Set $\overline{A(H)}$'' the complement of this area. The cc areas will become 
evident as synchronous regions during the numerical integrations
in section \ref{sec:results}.

\par Regarding the patient P(1) data,
 we call ``Set $A(P)$'' the area covered by the
destructed cc ribbons in Fig.~\ref{fig:01}b and  
``Set $\overline{A(P)}$'' the complement of it.
 The tumorous areas are clearly discerned at the top middle/right of the slice
 (see black arrow in Fig.~\ref{fig:01}b), where the 
white matter structure is disturbed as a result of the invading tumorous cells 
\cite{wang:2019}. 
Slice 19 has been chosen here as a displaying set because, on the one hand it is
in the middle of the brain and contains extended cc areas and on the other hand it has been
greatly affected by the tumor in P(1).

\par In the following sections we mostly present numerical results from
control subject H(1) and patient P(1), but similar results and conclusion are drawn from
the other two subjects. 
\medskip
\section{The Models}
\label{sec:models}

\par To explore the validity of our method and the consistency of 
our results across different neuronal models we  use here two
paradigmatic models for the numerical integration
of the potentials: the FitzHugh-Nagumo (Sec.~\ref{sec:FHN})
and the Leaky Integrate-and-Fire (Sec.~\ref{sec:LIF})
models.   The dynamics of the two models
and and their implementation in neuronal networks 
 are first described and 
at the end, in Sec.~\ref{sec:connectivity}, the
construction of the coupling matrix is presented, 
making use of the MRI-DTI data of the healthy and destructed
brain topology.
\medskip

\subsection{The FitzHugh-Nagumo Model}
\label{sec:FHN}

\par The single FHN  oscillator model was introduced in early 1960s
and consists of two coupled, nonlinear equations, one describing the evolution of a fast
activator potential variable $u(t)$ and the other a slower inhibitor variable $v(t)$ 
\cite{fitzhugh:1961,nagumo:1962}:
\begin{subequations}
\begin{align}
\label{eq01a}
\epsilon \frac{du}{dt}&=u-\frac{u^3}{3}-v\\
\label{eq01b}
\frac{dv}{dt}&=u+{\rm a} .
\end{align}
\label{eq01}
\end{subequations}
\noindent In Eq.~\eqref{eq01a}, $\epsilon$ is a parameter which accounts for the time-scale difference and is fixed in this study at $\epsilon=0.05$. The parameter $\rm a $ defines the dynamical behavior of Eqs.~\eqref{eq01}.
For  $|{\rm a} | <1$ the system contains an unstable fixed point, while for  $|\rm a |>1$ the fixed
point becomes stable.
In the region $|\rm a | <1$ the unstable fixed point gives rise to a limit cycle through a Hopf
bifurcation scenario. Throughout this study the parameter $\rm a$ is set to $\rm a =0.5$, in order
to assure the presence of the limit cycle.  For these parameter values the period
of the uncoupled FHN oscillator is 2.7 time units.
\par FHN oscillators will be used in this study
to model the dynamics in each voxel recorded by the MRI-DTI imaging technique. 
All voxels which
have non-zero density of neuron axons (non-white color in Fig.~\ref{fig:01}) are equipped with an 
FHN oscillator. 
\par The communication between the oscillators is dictated by the connectivity
matrix $\sigma_{ijk:lmn}$ which connects the oscillator
residing in voxel with 3D coordinates $(i,j,k)$ with the one in voxel $(l,m,n)$
(see details in Sec.~\ref{sec:connectivity}). 
\par Omelchenko et al., in Ref.~\cite{omelchenko:2013}, have shown
analytically and numerically that the coupled FHN model produces nontrivial partial synchronization
patterns when activator-inhibitor cross coupling terms are involved. An account of different 
functional interactions is given in \cite{pereda:2014}, while a variety of activator-inhibitor 
coupling schemes used in a biological framework is reviewed in Ref.~\cite{kiskowski:2004}.
Here, we use a coupling matrix with strong cross-coupling terms as proposed in Refs.~
\cite{omelchenko:2013,oomelchenko:2018}.
The coupled FHN dynamics is given by the following scheme:

\begin{widetext}
\begin{subequations}
\begin{align}
\epsilon \frac{du_{ijk}}{dt}&=u_{ijk}-\frac{u_{ijk}^3}{3}-v_{ijk}
- \sum_{(l,m,n)}\sigma_{lmn:ijk} \left[ b_{uu}(u_{lmn}-u_{ijk})+b_{uv}(v_{lmn}-v_{ijk})\right] \\
\frac{dv_{ijk}}{dt}&=u_{ijk}+{\rm a} - \sum_{(l,m,n)}\sigma_{lmn:ijk}
 \left[ b_{vu}(u_{lmn}-u_{ijk})+b_{vv}(v_{lmn}-v_{ijl})\right]
\end{align}
\label{eq02}
\end{subequations}
\end{widetext}
\par All FHN oscillators have the same parameters $\rm a$ and $\epsilon$ and 
the sums run over all voxels. 
The form of the coupling matrix
restricts the interaction between distant voxels and accounts for the boundary conditions, as will be discussed in 
Sec.~\ref{sec:connectivity}. 
A rotational matrix, $B$, is a simple way to parameterize the
possibility of diagonal coupling $(b_{uu},b_{vv})$ and 
activator-inhibitor cross coupling $(b_{uv},b_{vu})$ by a single parameter $\varphi$,
which is close to $\pi /2$ if cross-coupling is dominant \cite{omelchenko:2013}.
This coupling phase $\varphi$ is similar to the phase-lag parameter $\alpha$ of the
paradigmatic Kuramoto phase oscillator model, which is widely used to generically
describe coupled oscillator networks. The coupling phase is necessary for the
modeling of nontrivial partial synchronization patterns, as has been shown for the Kuramoto model 
\cite{oomelchenko:2010} and 
for the FHN model \cite{omelchenko:2013}.  
This rotational matrix $B$  was later used in 
Refs.~\cite{omelchenko:2015,schmidt:2017,ramlow:2019,mikhaylenko:2019} 
and has the form:

\begin{equation}
B=
\begin{pmatrix}
b_{uu} & b_{uv}\\
b_{vu} & b_{vv}
\end{pmatrix}
=\begin{pmatrix}
\cos{\varphi} & \sin{\varphi}\\
-\sin{\varphi} & \cos{\varphi}
\end{pmatrix}.
\label{eq03}
\end{equation}
 In the present study the value of the coupling phase 
$\varphi = \pi /2 -0.1$ is used. $\varphi $-values in a narrow band around $\varphi = \pi /2$ are known to lead to coexistence of coherent and incoherent domains in networks consisting of FHN oscillators
\cite{omelchenko:2013}.

\subsection{The Leaky Integrate-and-Fire model}
\label{sec:LIF}
The LIF model, introduced in 1907 by Louis Lapicque 
\cite{lapicque:1907a,lapicque:1907b}, 
describes the potential
activity of isolated neurons
via a single dynamical variable $u$.
The evolution equation of the variable $u$ comprises two stages, an integration
stage, Eq.~\eqref{eq04a}, and an abrupt resetting stage, Eq.~\eqref{eq04b} :

\begin{subequations}
\begin{align}
\label{eq04a}
& \frac{du}{dt}=\mu-u,\\
& \lim_{\varepsilon \to 0} u(t+\varepsilon ) \to u_\text{rest}, 
\>\>\> \text{when} \>\> u\ge u_\text{th}.
\label{eq04b}
\end{align}
\label{eq04}
\end{subequations}
\noindent The second of the two equations represents the resetting condition: 
If the membrane potential reaches a threshold $u_\text{th}<\mu$, it is reset to a resting potential $u_\text{rest}$.
Without loss of generality, we use the following parameter set: the parameter $\mu$
which represent the maximum achievable potential is set to $\mu =1$, the threshold potential
is set to $u_\text{th}=0.98$ and
the reset or resting potential is set to $u_\text{rest} = 0$. 
 For these parameter values the period
of the uncoupled LIF oscillator is $T_s=3.91$ time units.

\par During the integration stage, 
Eq.~\eqref{eq04a} behaves linearly and it can be solved analytically
to obtain the neuron potential as:
$u(t)=\mu -\left(\mu-u_0\right)e^{-t}$,
with the initial condition $u(0)=u_0$. This behavior is repeated after each resetting, while
the period $T_s$ of the single (uncoupled)
LIF oscillator depends on $u_\text{rest}$ and $u_\text{th}$, as
$T_s=\ln \left[\left(\mu -u_\text{rest}\right)/\left(\mu - u_\text{th}\right)\right]$.
Although the neurons are known to spend a refractory period after the resetting stage,
in this study the refractory period will be set to 0.

\par The coupling in the case of the LIF model follows the same scenario as in the
FHN model described in the previous section. The coupling is even simpler, since there
is only one variable and thus a rotational matrix is not needed.
Here, all voxels recorded in the MRI image which have non-zero density of neuron axons (non-
white color in Fig. 1) are equipped with a LIF oscillator.
The connectivity matrix between the oscillators is the same in the LIF and FHN models,
 $\sigma_{ijk:lmn}$, see Sec.~\ref{sec:connectivity}.
The LIF coupled network dynamics reads:
\begin{subequations}
\begin{align}
\label{eq05a} 
&\frac{du_{ijk}}{dt}=\mu-u_{ijk}-
\sum_{(l,m,n)} \sigma_{lmn:ijk}\left[u_{lmn} - u_{ijk}\right], \\
& \lim_{\varepsilon \to 0} u_{ijk}(t+\varepsilon ) \to u_\text{rest}, 
\>\>\> \text{when} \>\> u_{ijk}\ge u_\text{th}.
\label{eq05b}
\end{align}
\label{eq05}
\end{subequations}
\noindent All oscillators 
of the network have the same parameters, $u_\text{th}$, $u_\text{rest}$,
 $\mu $, while each one of them 
is reset independently, when reaching the threshold potential,
 $u_\text{th}$, common to all oscillators. 
In the present study we use values $u_\text{rest} < u_\text{th} < \mu $
and, therefore, the oscillators remain supra-threshold; their
potentials take values between $u_\text{rest}$  and  $u_\text{th}$.
Regarding initial conditions, all
oscillators start from $u_{ijk}(t=0)$ values randomly distributed between 
$0\le u_{ijk}(t=0) < u_\text{th}$.
\par  We stress that in the present study both models, LIF and FHN, operate in the
parameter regions which supports regular, periodic, self-sustained 
oscillations in the absence of any coupling.

\subsection{Connectivity}
\label{sec:connectivity}

Two types of connectivity matrices are used: $\sigma^H$ is the generic name
for matrices extracted from the MRI-DTI data
of healthy controls and $\sigma^P$ corresponds to the matrices extracted from the 
 data
of patients. 

\par Generally, the MRI-DTI data (fractional anisotropy values) give the
local density $I(i,j,k)$ of neuron axons present in the voxel centered around the coordinates
$(i,j,k)$. The element of the connectivity matrix
($\sigma^H$ for the healthy subjects and $\sigma^P$
for the patients) which links voxels $(l,m,n)$ and $(i,j,k)$, is computed as:
\begin{widetext}
\begin{eqnarray}
\sigma_{ijk:lmn}=\begin{cases}
\dfrac{h }{N_R}\dfrac{I(i,j,k)\>\> I(l,m,n)} 
{ \displaystyle{\sum_{\rm all \> voxel-pairs} I(p1,p2,p3)\>\> I(q1,q2,q3)}}, & \text{where} \>\>\>
 (i-l)^2\le R^2, \>\>\> (j-m)^2 \le R^2 \>\>\> \text{and} \>\>\> n=k\pm 1 \\
0, & \text{otherwise}.
\end{cases}
\label{eq06}
\end{eqnarray}
\end{widetext}
\par The product in the numerator of  formula \eqref{eq06} ensures that only if there is 
nonzero white matter intensity in both $(l,m,n)$ and $(i,j,k)$
 voxels interconnection between the voxels
takes place.  
The sum over all pairs $(p1,p2,p3)$ and $(q1,q2,q3)$ in the denominator is set
for normalization purposes.
The network exchanges are limited to distances less than
a given $R$ in the x-y plane (slice)  and each slice is
 linked only to the layers
above and below it, $n=k+1, \> k-1$.
 In this way the brain is considered as a
multilayer network 
\cite{majhi:2016,schoell:2019} consisting of $n_s$ layers, 
where each node $(i,j,k)$ interacts with nodes
$(i-R,j,k)$, $(i-R+1,j,k)$ ... $(i+R,j,k)$ and $(i,j-R,k)$, $(i,j-R+1,k)$ ... $(i,j+R,k)$ 
in the same slice,
and with nodes $(i,j,k+1)$ and $(i,j,k-1)$ in the perpendicular direction.
The total number of links of 
each unit, denoted by $N_R$, is at most  $(2 R +1)^2+1$, 
i.e., $(2 R +1)^2-1$ in its layer, plus 2 nodes above and
below, in the axial direction.  The factor $N_R$  counts the number of interacting 
neighbors and is set in the denominator for
normalization purposes
 and ensures proportional contribution of each node in the
sum. Note that the nodes which belong to the borders of the structure have less 
than $(2 R +1)^2+1$ neighbors. For these nodes the $\sigma_{ijk:lmn}$ elements are divided by
the number of existing neighbors. Overall, the coupling terms
 are  multiplied (leveled) by a factor $h$ to modulate the intensity of the interactions. 

\par  We recall from Sec.~\ref{sec:data} that the connectivity matrices used 
in this study are 3D with planes of size $256 \times 256$ and are composed 
by $40 \le n_s \le 44$ 
such planes.
The coupling range $R=25$ is used, which is an intermediate
coupling between local interactions $R=1$ and all-to-all coupling. This intermediate $R$-value
 accounts both
for nearest neighbor interactions which are attributed to electrical exchanges and long distance
coupling which is mediated by neurotransmitters. Intermediate coupling ranges
 between global (all-to-all) and nearest-neighbor connectivities are observed in many natural systems. Previous studies \cite{omelchenko:2013,omelchenko:2015} 
have shown that there exists a quite wide intermediate 
range where the partially synchronized patterns, chimera states, can be observed. 
Moreover, these patterns are robust to the inhomogeneity of the network nodes or 
slight changes in the topology. We select an exemplary value for $R$ in our numerical 
simulations, but other values of $R$ in the range $10 \le R \le 60$
give qualitatively consistent results.

\par To avoid misunderstanding between structural and functional
connectivity matrices, we stress here that the matrices
extracted using Eq.~\ref{eq06}, are
structural, weighted connectivity matrices which are directly recorded by 
MRI scanners
using the MRI-DTI technique.
These matrices are not directly related with the functional connectivity 
matrices used in the literature. For the formation of the functional connectivity matrices the 
brain is lumped in 60-90 functional cortical regions and the exchanges between these regions
are established using mostly EEG techniques and more recently MRI techniques 
\cite{hagmann:2008,castro:2020,carboni:2019,griffa:2019}. Instead, in this work the spatial
arrangement of the voxels is taken into account as directly represented by the MRI images.
Nevertheless, the use of a dynamic model (FHN or LIF) in the network simulations
 allows us to extract functional synchronization patterns. 

\par Following the above rules the full spatial connectivity matrices are constructed
 for the healthy and patient subjects 
and these matrices are used in Eqs.~\eqref{eq02}
and \eqref{eq05} for the integration of the potentials.

\section{Results}
\label{sec:results}

In this section we present the results of numerical integration of the FHN equations and the LIF 
equations where each voxel is considered as an oscillatory unit with connectivity matrices obtained
 as discussed in the previous section, \ref{sec:connectivity}. Firstly, the synchronization
properties of the healthy control subject will be presented (section \ref{results:healthy-FHN}) 
followed by the patient results (section \ref{results:anomalous-FHN}).
To test the validity of the results, the same MRI-DTI data
 of the control subject and the patient
will be used in sections \ref{results:healthy-LIF}
and \ref{results:anomalous-LIF}, respectively,
but the LIF dynamics will be used for the numerical integration. 
Although the dynamical
schemes employed are entirely different, we demonstrate that the main qualitative
synchronization patterns persist and the pacemaker effect  which differentiates
 the control subjects from the
patients is evident using both dynamical systems.
\par 
As working parameters the following sets were used: For the FHN system $\epsilon = 0.05$, 
$\rm a = 0.5$,  $\varphi = \pi - 0.1$, $h = 0.1$;
 for the LIF model,  $\mu = 1.0$, $u_{\rm rest} = 0.0$,
$u_{\rm th} = 0.98$, $h = 0.6$ and the coupling parameter is $R = 25$ in all cases. 
In the simulations both the Runge-Kutta 4th order method 
with integration step $10^{-2}$ and the Euler method with integration step $10^{-3}$ and 
$10^{-4}$ were used.

\subsection{FHN Results on Healthy Brain Structure}
\label{results:healthy-FHN}

FHN oscillators
 were placed on all voxels containing neuron axons. As an example, in plane (slice) 19
shown in Fig.~\ref{fig:01}a identical FHN 
units were placed on all non-white cells and
similarly for all other planes above and below this. The FHN 
units were coupled nonlocally
along the x-y plane and locally along the z-axis constituting a multilayer network
composed of $n_s=44$ layers along the z-axis,
as described in Sec.~\ref{sec:connectivity} and Eq.~\eqref{eq06}. The connectivity matrix, 
$\sigma ^H$, used here was extracted from  the
  data of the healthy subject H(1) following the discussions
 in sections~\ref{sec:data} and ~\ref{sec:connectivity}.
Numerical integration of the corresponding Eqs.~\eqref{eq02} and \eqref{eq03}
 was carried out for 2000 time units
starting from random initial conditions (potentials).

\begin{figure}[h]
\includegraphics[width=0.48\textwidth,angle=0.0]{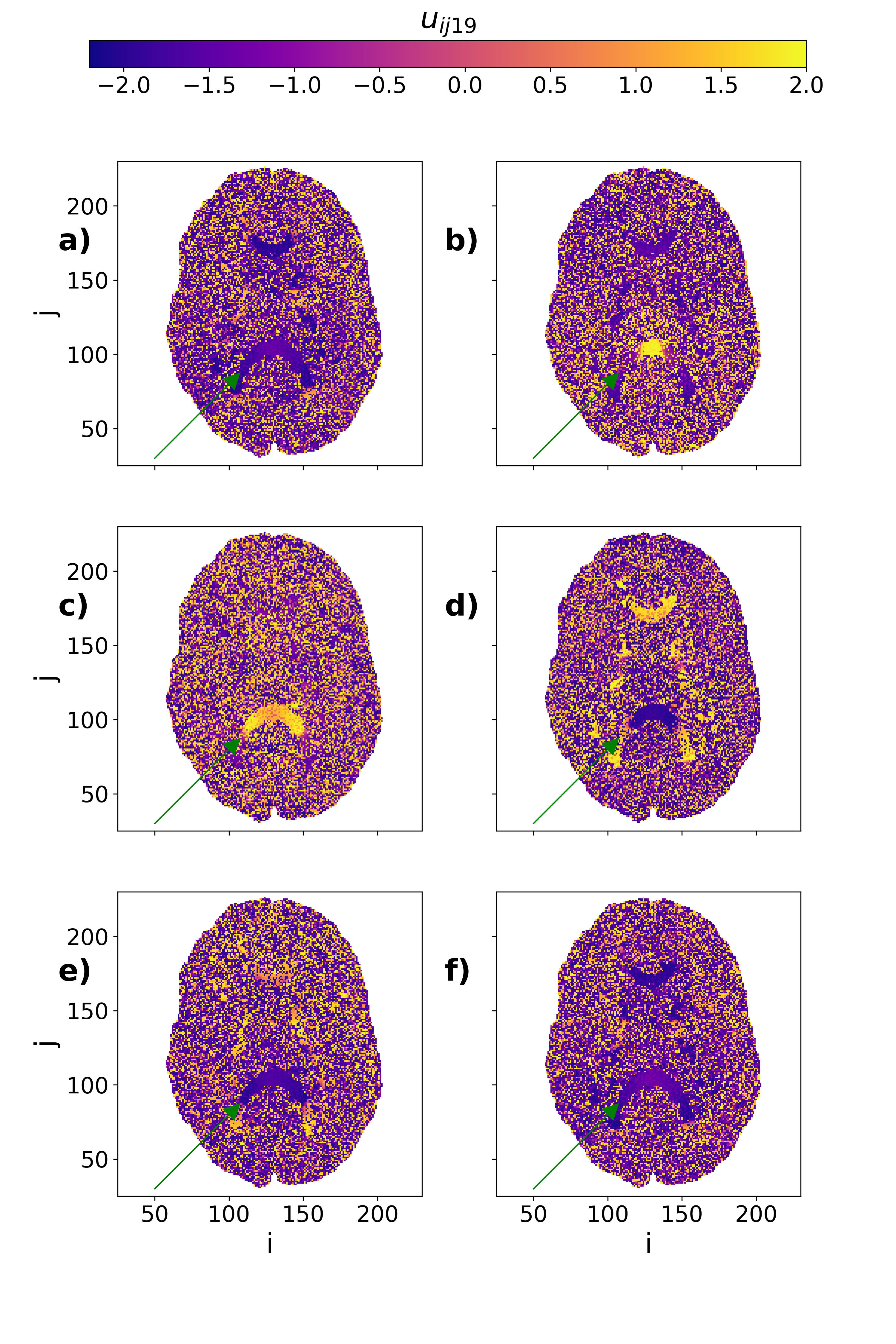}
\caption{\label{fig:02} Color-coded typical potential profiles (snapshots)
of slice 19 of healthy control H(1) 
 using FHN dynamics. The profiles
are recorded at 
  a) 847.0, b) 847.8,
c) 848.3, d) 849.0, e) 849.4 and f) 849.8 time units within one complete period.
The green arrows point to the cc areas.
Simulations start from random
initial potentials. Other parameters are: 
$h=0.1$, $\varphi =\pi - 0.1$, $\epsilon =0.05$, $\rm a =0.5$ and $R=25$.  
}
\end{figure}

\par After a transient period the network stabilizes in 
a state where the  cc ribbon areas,
set $A(H)_{\rm FHN}$, constitute the  
 coherent parts, while the rest of the brain, the complement set $\overline{A(H)_{\rm FHN}}$ is
incoherent.
Representative color-coded potential results on slice 19 are presented in Fig.~\ref{fig:02}. 
 These are typical 
steady state snapshots showing
the different phases of the slice. The green arrows point to the cc areas.
  The six representative snapshots are all recorded
within an interval of 3 time units, which is the maximum 
period observed in the system. In particular: panel a) is characterized by low, common potential
values in all cc areas and by mixed state of the complement set; 
in panel b) the potential in the middle of the lowest cc area starts increasing (becoming yellow)
while the complement set keeps in the incoherent state; in panel c) the yellow color (high potential 
values) invades the lower cc area, while the higher cc area also increases its potentials and becomes
hidden within the incoherent set; in panel d) the potentials in the lower cc area start increasing
(becoming dark blue) from the center,  while the upper cc area becomes again visible having simultaneously acquired the high potential values (yellow colors); in panel e) the low potentials (dark blue colors)
start invading the lower cc area while the potentials in the upper cc area take intermediate values
(orange colors) and in panel f) all cc areas acquire the lowest (dark blue)
potential values. The f) state is equivalent to
state a) and are considered as the starting point of a new period.
Note that in regards to the cc areas, panels b) and d) are in opposite phases of each other.

\par 
From the results in Fig.~\ref{fig:02} we note that while in the cc areas the potentials
are identical or follow closely their neighboring values,  
 the complement sets are always in the incoherent state.
This behavior can be considered
as a ``chimera-like'' state.
(Note that the  cc areas correspond to brain regions of high density, where the connectivity
between the two hemispheres, left $\leftrightarrow$ right, dominates,
see Fig.~\ref{fig:01}a.)  We recall here that the term ``chimera state'' is reserved to networks where synchronous
and asynchronous regions coexist under the condition that  {\it all nodes
are identical and identically linked}. In the present case, all oscillators are identical but the
linking is not homogeneous since it is recorded from the MRI-DTI data of a real healthy brain. For this
reason, as explained in the Introduction, the term ``chimera-like state'' is used. 

\begin{figure}[h]
\includegraphics[width=0.48\textwidth,angle=0.0] {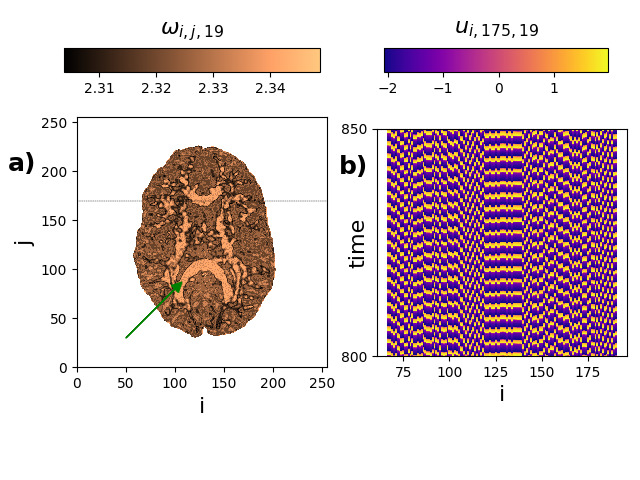}
\caption{\label{fig:03} 
  a) Mean phase velocity profile of slice 19 of healthy subject H(1) using FHN dynamics. b) 
Spacetime plot corresponding to the cut for $j=175$ indicated in (a) with the grey line.
The green arrow points to the cc area. 
The average in (a) was taken over 2000 time units. 
 Other parameters as in Fig. \ref{fig:02}.
}
\end{figure}

\par The chimeric nature of the coupled system
 can be further verified using the mean phase velocity profile. The mean phase velocity $\omega_{ijk}$
accounts for the number of cycles $c_{ijk}$ that oscillator at position $ (i,j,k)$ has performed during
a certain time interval $\Delta T$, and is defined as:
\begin{eqnarray}
\omega_{ijk}=2\pi \frac{c_{ijk}}{\Delta T}=2\pi f_{ijk}\>\>,
\label{eq07}
\end{eqnarray}
 where $f_{ijk}$ is the frequency of the oscillator. 
Because the definitions of frequency and mean phase
velocity differ by just a factor $2\pi$, in the following the two expressions are used interchangeably.
\par The cc ribbon sets $A(H)_{\rm FHN}$ of slice 19 demonstrate common mean phase velocity
and they clearly correspond to the cc ribbon sets $A(H)$ of the MRI-DTI data, compare 
Figs.~\ref{fig:01}a and ~\ref{fig:03}a (the green arrows point always to the cc areas). 
The mean phase velocity over 
the ribbon set is constant and is consistently higher than in the rest of the structure.
Returning to Fig.~\ref{fig:02}, on the lower cc ribbon area we can discern  color-grading  
which corresponds to traveling waves in the corpus callosum. 
These are more evident in the videos included in the Supplementary Material (Multimedia view). The
traveling waves disappear in the complement set $\overline{A(H)_{\rm FHN}}$ and they are held
responsible for the higher mean phase velocities observed in the cc areas.

\par To elucidate further the nature of oscillations in the healthy brain, 
in Fig.~\ref{fig:03}b we present the spacetime plot of the potential along a line crossing
slice 19 at position $j=175$, see grey horizontal line in Fig.~\ref{fig:03}a. This line was chosen
because on the one hand it crosses both $A(H)_{\rm FHN}$ and $ \overline{A(H)_{\rm FHN}}$
sets in the healthy brain and, on the other hand, it crosses at the same level
the tumorous regions in the patient P(1) brain (see Sec.~\ref{results:anomalous-FHN} and Fig.~\ref{fig:05},
later on).
In Fig.~\ref{fig:03}b, it is possible to see that the synchronous cc regions coexist with the
asynchronous
complement  $ \overline{A(H)_{\rm FHN}}$, forming the chimera-like state.

\subsection{FHN Results on Anomalous Brain Structure}
\label{results:anomalous-FHN}

\begin{figure}[t]
\includegraphics[width=0.48\textwidth,angle=0.0] {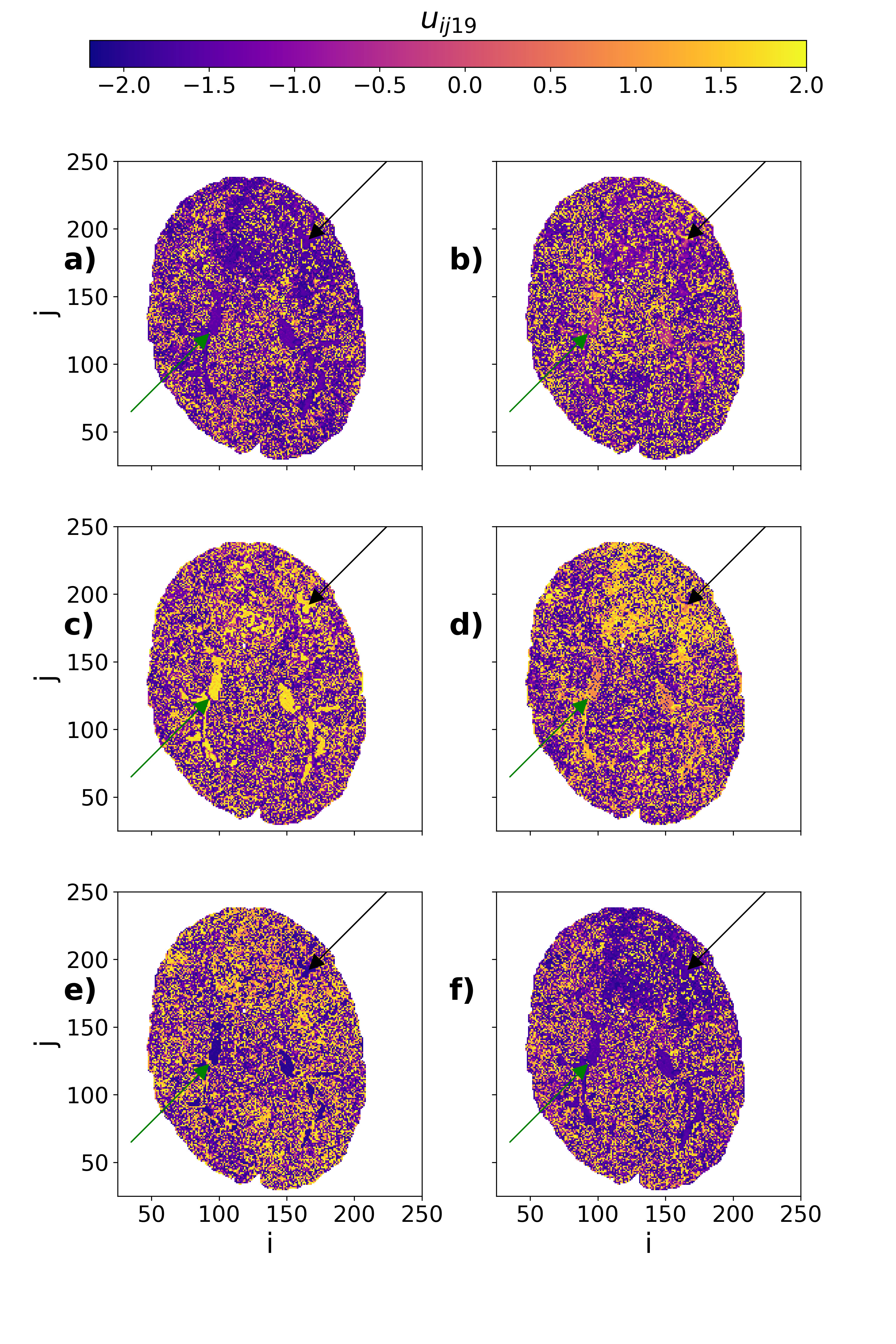}
\caption{\label{fig:04} Color-coded typical potential profiles 
of slice 19 of patient P(1) with destructed brain structure using FHN dynamics. 
The profiles
are recorded  at 
  a) 382.9, b) 383.5,
c) 383.9, d) 384.4, e) 384.7 and f) 385.4  time units. 
The green arrows point to cc areas and black arrows point 
to the tumorous area.
Simulations start from random
initial potentials. 
 Other parameters as in Fig. \ref{fig:02}.
}
\end{figure}

Exactly the same numerical simulation process was applied here, as in the previous section~\ref{results:healthy-FHN},
the only difference being that now the connectivity matrix, $\sigma^P$,
 extracted from the patient P(1) was used. 
Namely, identical FHN oscillators were placed on all voxels and
 these where coupled nonlocally
along the x-y plane and locally along the z-axis.
Numerical integration of Eqs.~\eqref{eq02} using the connectivity matrix $\sigma ^P$
 were carried out for 2000 time units
starting from random initial potentials. The color-coded potential
results on slice 19 are presented in Fig.~\ref{fig:04} at six different instances
within the maximum period of oscillations.
At all instances, the cc structure (see green arrows)
is presented damaged, as also shown in the MRI-DTI data 
(Fig.~\ref{fig:01}b).
\par A new observation is the ``pacemaker effect'' which is more clearly visible in
the supplementary video (Multimedia view)
and also appears in the simulations using the LIF model in section
~\ref{results:anomalous-LIF}.
 Namely, potential waves are initiated in the tumorous region 
(see black arrows in Fig.~\ref{fig:04}), they 
propagate through the white matter and they disappear at the borders of the brain structure before
a new potential wave restarts at the tumorous region. 
More precisely, 
 in panel  \ref{fig:04}a) the regions around the tumorous areas demonstrate low (dark blue)
potentials similar to the destructed cc areas; in panel b) the potentials in the tumorous area increase 
toward orange-yellow values and so do the cc areas; in panel c) the tumorous areas acquire maximum
potential values; in d) the maximum potential values propagate away from the tumor in the 
cc-complement set while the cc set has intermediate (orange colored)
values; in e) the potentials in the tumorous regions start decreasing toward dark blue values
and in f) the dark blue values propagate covering the tumorous region, while the cc regions
also acquire the lowest potential values. Panel f)  corresponds to panel a) which was
considered as the starting stage of a period.
The gradual propagation of the yellow color from the tumorous areas to the rest of the brain
(sequence of panels b) $\to$ c) $\to$ d) )
illustrates the pacemaker effect which is centered in the
lesion and can be used potentially to identify the position of the tumor.

\begin{figure}[h]
\includegraphics[width=0.48\textwidth,angle=0.0]{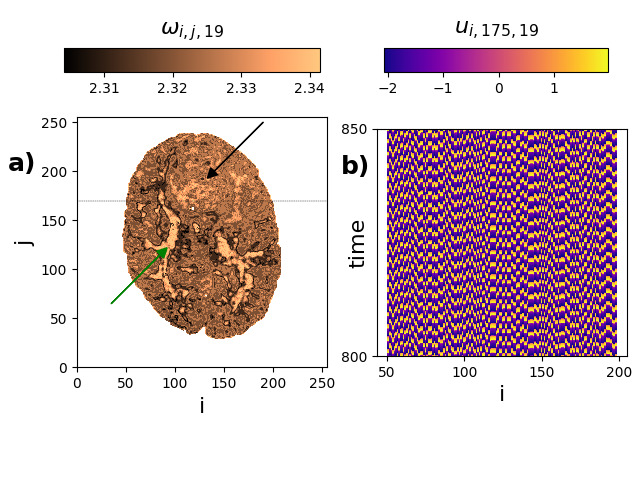}
\caption{\label{fig:05} 
  a) Mean phase velocity profile of slice 19 of the subject P(1) 
with destructed brain structure using FHN dynamics. 
 b) Spacetime plot corresponding to the cut for $j=175$ indicated in (a) with the grey line
which crosses the lesion. 
 The average in (a) was taken over 2000 time units. The green arrow points
 to cc areas and the black arrow
to the lesion area.
 Other parameters as in Fig. \ref{fig:02}.
}
\end{figure}

\par The associated mean phase velocity profile in Fig.~\ref{fig:05}a
indicates clearly the positions of the destructed ribbon structure (see green arrow),
where high $\omega$ values are observed. In the same image  
the tumorous area is shown, marked by the black arrow, where also relatively 
high mean phase velocities are recorded. 
Comparing Figs.~\ref{fig:03}a and \ref{fig:05}a,  
it is evident that high  $\omega$'s dominate in the position where the potential waves are
initiated in the lesion area.
\par Similarly to the healthy case, Fig.~\ref{fig:03}b,
in Fig.~\ref{fig:05}b  the spacetime plot of the potential along a line crossing 
slice 19 at position $j=175$ is presented (see grey horizontal line in Fig.~\ref{fig:05}a).
 This line now  crosses the tumorous regions.
In Fig.~\ref{fig:05}b,  it is not possible to discern any synchronous region. The tumor has destructed
the tissue structure and the chimera-like profile observed in the healthy brain, Fig.~\ref{fig:03}b,
is no longer manifested. This difference can be used as an indicator of malignancy
and can be critical in cases where the tumor is small, not visible by eye, but it can destroy the
chimera-like patterns and  induce brain waves
through the pacemaker effect. 

\par Although the results presented in the above two sections depict layer 19
for the healthy and the patient subjects, qualitatively similar results and
conclusions are drawn from all other layers in the network.
In the next two sections,  the FHN dynamics is replaced by the LIF model dynamics 
and it is shown that the pacemaker effect is robust across
models (is not model dependent) since it comes as a result of the brain structure anomalies.

\subsection{LIF Results on Healthy Brain Structure}
\label{results:healthy-LIF}

\begin{figure}[h]
\includegraphics[width=0.48\textwidth,angle=0.0]{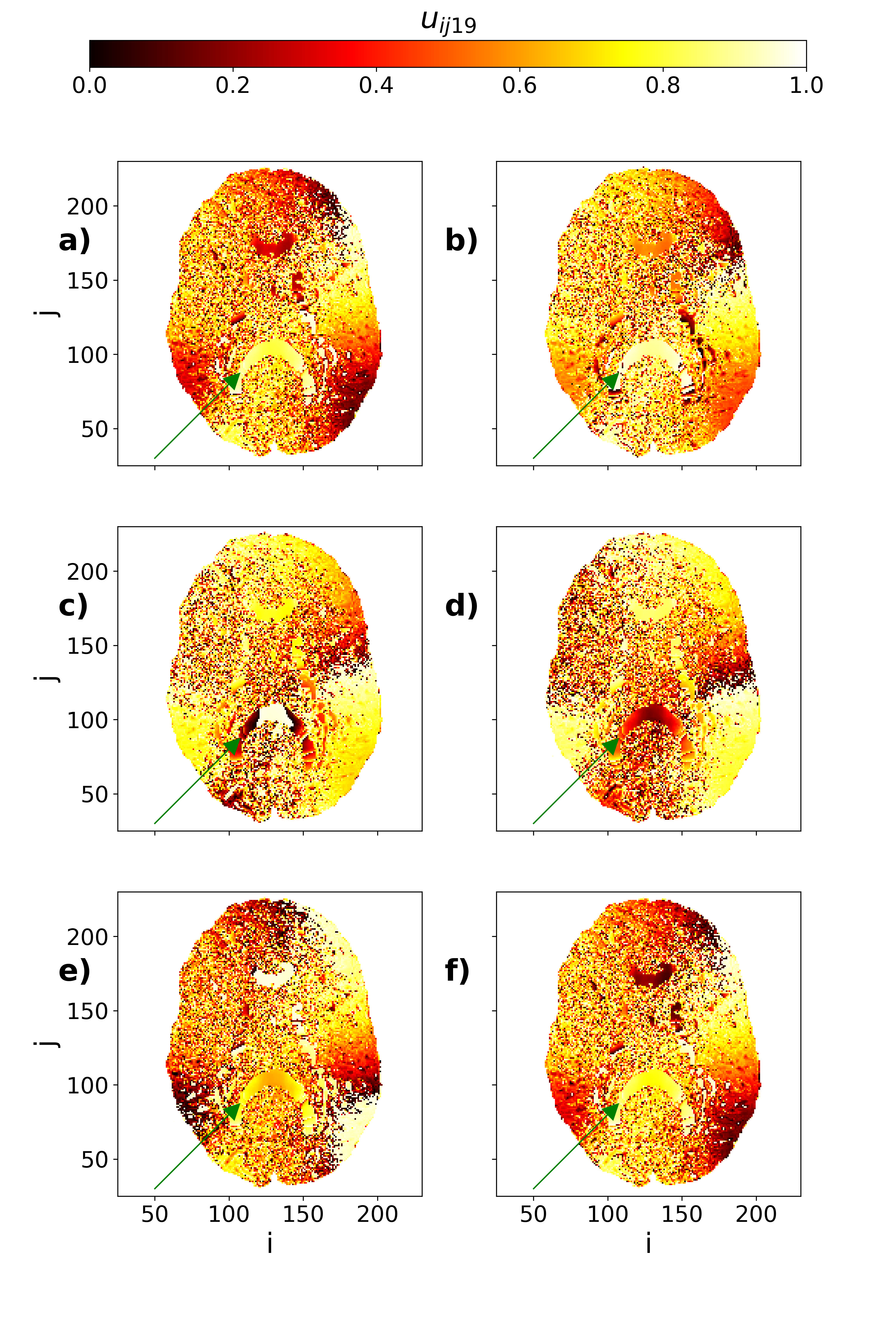}
\caption{\label{fig:06} Color-coded potential profile of slice 19 of healthy control 
H(1) using LIF dynamics. The profiles
are recorded  at 
   a) 846.7, b) 847.2,
c) 847.8, d) 848.2, e) 849.0 and f) 849.4  time units  (for comparison with
healthy brain network dynamics in section~\ref{results:healthy-FHN} ). 
Simulations start from random
initial potentials. Other parameters are: $h =0.6$, $\mu =1.0$, $u_{\rm rest}=0.0$ and $u_{\rm th}
=0.98$. Green arrows as in Fig.~\ref{fig:02}.
}
\end{figure}

In this section, the LIF scheme, Eq.~(\ref{eq05}), is employed to simulate the dynamics on all nodes of the 
healthy subject H(1). The
connectivity matrix used is $\sigma^{H}$, the same one which was used in section~\ref{results:healthy-FHN}.
All oscillators
 start from random initial potentials in the range [0,$u_{\rm th}$].
Figure ~\ref{fig:06} depicts six color-coded potential profiles of representative slice 19, 
centrally located
on the vertical axis of the head. These typical profiles represent the various phases that the cc and 
complement domains undergo within one period of LIF oscillations.

\begin{figure}[h]
\includegraphics[width=0.48\textwidth,angle=0.0]{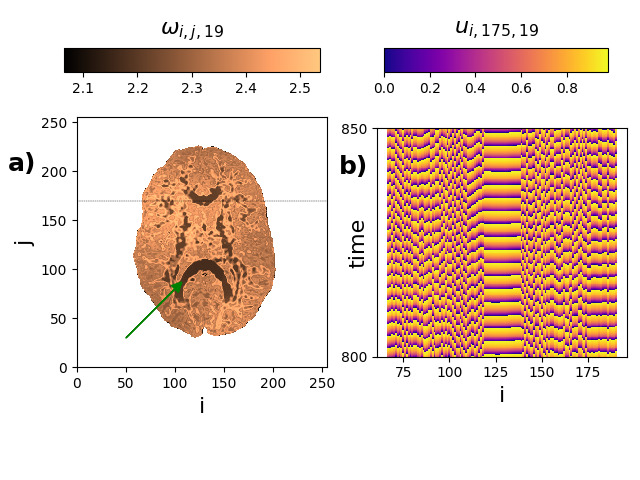}
\caption{\label{fig:07} 
  a) Mean phase velocity profile of slice 19 of healthy subject H(1) using LIF dynamics. b) 
 b) Spacetime plot corresponding to the cut for $j=175$ indicated in (a) with the grey line.
The average in (a) was taken over 2000 time units. Green arrow as in Fig.~\ref{fig:03}.
Other parameters as in Fig. \ref{fig:06}. 
}
\end{figure}
\par Clearly, two different regions can be identified in all six panels: 
a) A set $A(H)_{\rm LIF}$ consisting of the isolated cc segments which present phase coherence 
(green arrows in Fig.~\ref{fig:06})
and b) the complement $ \overline{A(H)_{\rm LIF}}$
of this set where the phases are incoherent.  The sets $A(H)_{\rm LIF}$ and $ \overline{A(H)_{\rm LIF}}$   
overlap with the corresponding sets  $A(H)$ and $ \overline{A(H)}$ of the MRI-DTI data
 of the healthy subject H(1).
 A simple comparison between LIF and
FHN numerical results indicates that the sets $A(H)_{\rm LIF}$ and $ \overline{A(H)_{\rm LIF}}$ 
(see Fig.~\ref{fig:06}) present the same characteristics 
with the corresponding sets  $A(H)_{\rm FHN}$ and $ \overline{A(H)_{\rm FHN}}$ resulting from FHN simulations
(see Fig.~\ref{fig:02}). 

\par The panels in Fig.~\ref{fig:06} demonstrate the LIF activity
within one period of oscillations. For clarity we present here the evolution
within the cc regions which in the LIF model keep a constant phase difference.
In panel a) the lower cc region is characterized by high (yellow) potential values while the
top cc region has intermediate (red) values; in panel b) the potentials increase in the lower cc region
becoming almost white (value $u_{ijk}\sim 1.0$), 
with a similar increase in the top cc area toward orange color;
in panel c) the center of the lower cc region is still in the range of higher ($u_{ijk}\sim 1.0$) values
while its border voxels
 have dropped to zero (black and dark red color) due to the resetting
condition, Eq.~\ref{eq05b}, and the top cc region potentials increase toward yellow values;
in panel d) the dark and red regions (low potentials) have now invaded the lower cc area, while in the
upper one the potentials increase further; in panel e) the potentials in the lower cc area increase
toward orange-yellow values while in the top cc area the potentials have reached their maximum
values  ($u_{ijk}\sim 1.0$) and in panel f) the lower cc area potentials increase further 
(yellow colors) while in the upper cc area  resetting occurs and the potential values drop to
dark and red values. This last panel f) is similar to panel a) which is taken as the starting 
point of a new period. 
The related video provided in the Supplementary Material shows that a) the cc areas 
behave coherently demonstrating
 a consistent phase difference between the upper and lower regions 
 and b) the activity in the complement
is in the form of irregular potential
waves which appear and disappear in the periphery of the brain (Multimedia view). 

\par The mean phase velocity profile of the same plane demonstrates the coexistence of coherent and incoherent
domains, see Fig.~\ref{fig:07}a. 
The cc ribbons of high white matter  density, set ${A(H)_{\rm LIF}}$, appear here undestructed,
their $\omega$-values present consistently lower values than the surrounding complement set
 and depict the same structure as set $A(H)$ of the MRI-DTI data, Fig.~\ref{fig:01}a.

\par Figure~\ref{fig:07}b depicts the spacetime plot of a linear cut along the slice 19, at position $j=175$,
as designated in Fig.~\ref{fig:07}a with the grey line.
At the position where this cut crosses the cc ribbons,  $120<i<135$, a coherent domain is noted in the
spacetime plot. This finding, using the LIF dynamics, agrees with 
similar findings depicted in Fig.~\ref{fig:03}
and Sec.~\ref{results:healthy-FHN}, where the FHN dynamics was employed. The fact that both FHN and LIF
dynamics confirm  the presence of chimera-like patterns in the healthy brain is an indication that 
this effect is
 model independent
and is caused by the interplay between the complex connectivity and the 
nonlinear dynamics added in the  network.

\subsection{LIF Results on Anomalous Brain Structure}
\label{results:anomalous-LIF}
\begin{figure}[h]
\includegraphics[width=0.48\textwidth,angle=0.0]{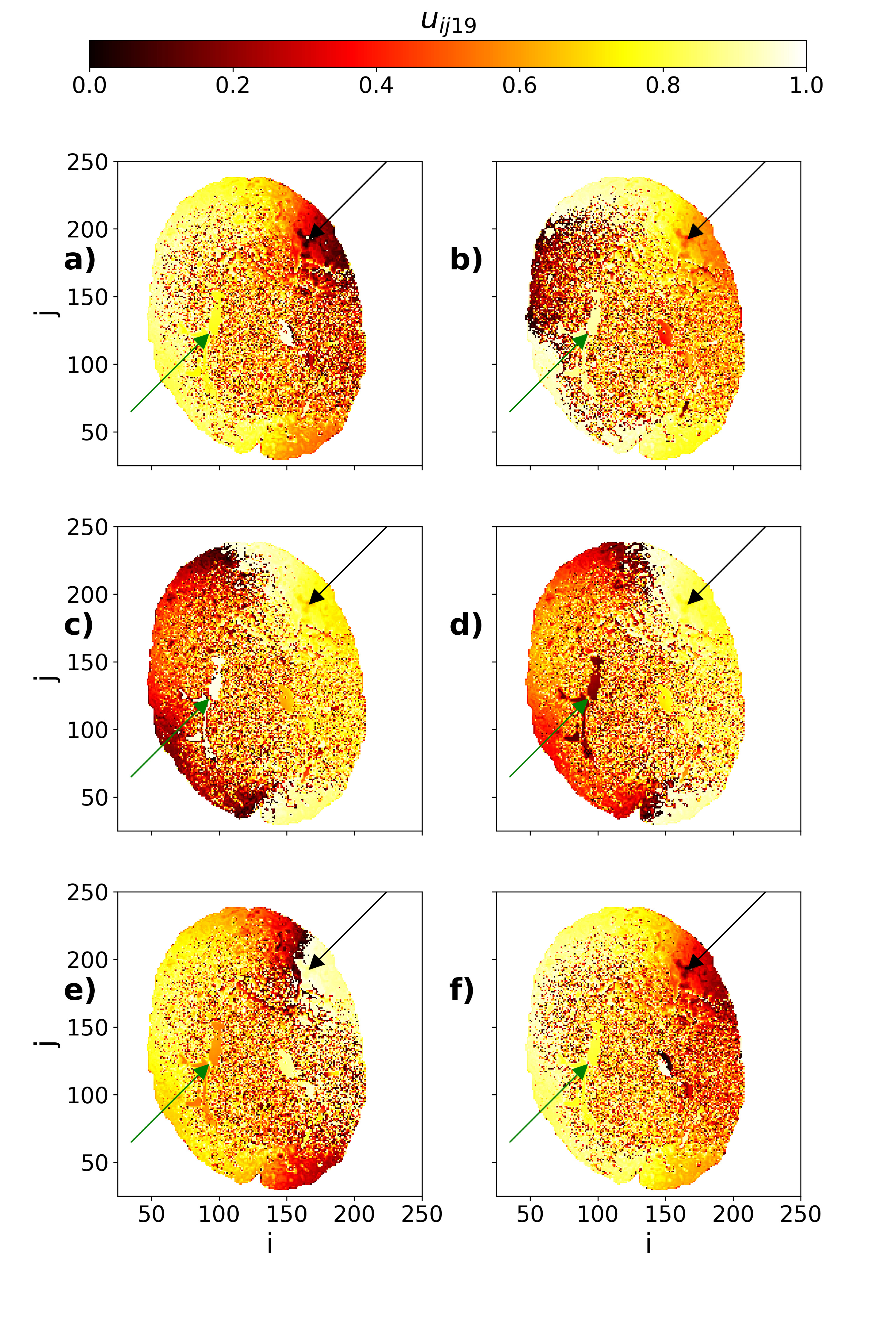}
\caption{\label{fig:08} Color-coded potential profiles of slice 19 of subject P(1)
with destructed brain structure using LIF dynamics. 
The six profiles
are recorded  at 
  a) 350.7, b) 351.4,
c) 351.9,  d) 352.2, e) 352.9 and f) 353.6 time units.
Simulations start from random
initial potentials.  Green and black  arrows as in Fig.~\ref{fig:04}.
Other parameters as in Fig. \ref{fig:06}.
}
\end{figure}

\par The numerical integration results on anomalous brain structure using the LIF dynamics 
are in qualitative agreement with the ones obtained using the FHN dynamics. 
To illustrate this, in Fig.~\ref{fig:08} we present the 
 representative phases characteristic of
the potential profiles: in panel a) the tumorous region is in the lowest potential
values (black and red colors) while the destructed cc area marked by the
green arrow is characterized by higher (yellow) potentials;
in panel b) the potentials in the tumorous region increase toward orange values while the
cc area potentials increase further toward lighter yellow; in panel c) we observe further
increase in the tumorous region toward yellow, while the cc area has reached maximum (white)
potentials; in panel d) the tumorous region potentials increase toward lighter yellow while
the cc pointed area potentials were reset to low values (dark, red); in panel e) the tumorous
region develops maximum (white) potentials while the $u$-values increase in the cc pointed
ribbon toward medium (orange) values and in panel f) the potentials are reset in the tumorous
region (dark, red colors) and the $u$-values in the cc region increase to high (yellow) values.
Note that panels a) and f) correspond to equivalent states and are selected as
 the starting points of a new period. Here the remaining disjoint pieces of the cc areas
present a phase shift in their potential values. For example, in panel a) the right cc area
is white-colored, higher than the left one (yellow colored). In panel b) the right cc area
was reset to low values while the left one increases further toward maximum potentials, and
so on.
\par The six panels of Fig.~\ref{fig:08} are indicative of the pacemaker effect
induced by the tumorous region which can be also seen in the following way: In panel a)
a yellow colored region appears in the left-top periphery of the brain. In panel b) it gives
rise, after resetting, to black and red regions. In panel c) the black-red region splits 
in two which propagate peripherally toward the tumorous regions. 
In panel d) the  black and red regions (low potentials)
propagate further toward the tumor, while behind them, to the left, the yellow region restarts 
developing. In panel e) the red regions approach further the tumorous area while behind them the
yellow region extends covering the entire left semiplane. In panel f) the red areas have covered/invaded
the tumorous region returning to a state similar to a). The effect is more clear 
in the corresponding video provided in the Supplementary Material (Multimedia view). After a transient
period of about 200 time units, the waves created in
the periphery of the brain collide in the tumorous region and disappear there. In this case, the
pacemaker acts as an absorber of the potential waves propagating circularly in the brain periphery.
This behavior is similar to the pacemaker effect that
was also observed in the FHN model with the difference that here the pacemaker acts as an absorber,
or a pacemaker with negative propagation.
\par  The cc ribbon structure, set $A(P)_{\rm LIF}$
 of high white matter density is also destructed in the LIF simulations, in agreement
with the MRI-DTI data (Fig.~\ref{fig:01}b) and the FHN results (Fig.~\ref{fig:04}). 
The region where the potential waves collapse is identified as the blue region in 
Fig.~\ref{fig:01}b, where the lesion is located. 

\par
\begin{figure}[h]
\includegraphics[width=0.48\textwidth,angle=0.0]{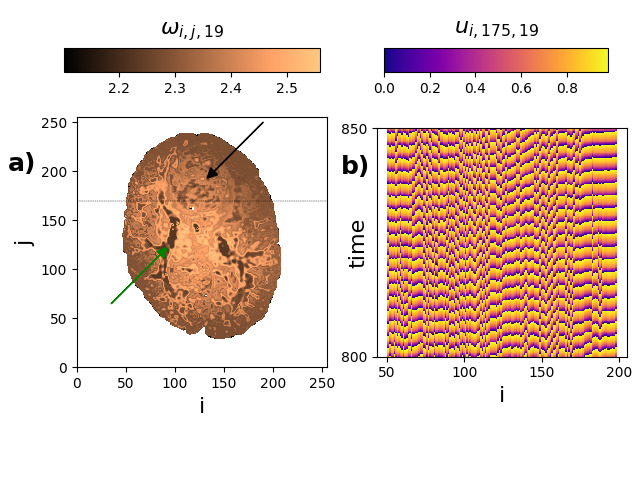}
\caption{\label{fig:09} 
  a) Mean phase velocity profile of  slice 19 of subject P(1)
with destructed brain structure using LIF dynamics.
 b) Spacetime plot corresponding to the cut for $j=175$ indicated in (a) with the grey line
which crosses the tumorous region. The average in (a) was taken over 2000 time units. 
Green and black  arrows as in Fig.~\ref{fig:05}.
Other parameters as in Fig. \ref{fig:06}.
}
\end{figure}

 The corresponding mean phase velocity profile results, Fig.~\ref{fig:09}a, 
conform with the above discussion. The tumorous
area indicated by the black  arrow
shows lower mean phase velocity than the center of the brain where the structure is less affected
by the pathology. Figure~\ref{fig:09}b depicts the spacetime plot of a linear cut along the slice 19 at position $j=175$ (designated in Fig.~\ref{fig:09}a with the grey line),
which crosses the tumorous region. Contrary to the case of healthy brain, 
Fig.~\ref{fig:07}b,  the spacetime plot here does not contain a coherent domain,
 due to the destruction of the cc ribbon areas.
Similar behavior was also noted earlier, in Fig.~\ref{fig:05}
and Sec.~\ref{results:anomalous-FHN}, where the FHN dynamics was used.


The qualitative agreement in the pacemaker effect produced in the patient using both LIF and FHN 
dynamics comes as a result of the destruction of the brain structure and is not an artefact of the 
dynamics, since it appears independently in the two models when the tumorous region is introduced through
the connectivity matrix. The anomaly in the present structure has a relatively large extension, as it
was the case for this particular patient. 
It would be interesting to explore if similar pacemaker effects will be
produced when smaller tumors are considered.

\par Overall, both FHN and LIF dynamical schemes
confirm  the presence of chimera-like patterns in the healthy brain 
while incoherent dynamics and pacemaker effects are observed in the destructed, tumorous brains.
These phenomena seem to be generic and not model dependent or artefacts of a particular model.

\section{Spectral Analysis of Healthy and Anomalous Dynamics}
\label{sec:spectral}
In this section a first attempt is made to test whether the presence of the tumor
affects the Fourier spectra of the neuronal oscillations locally, 
in the lesion areas. For this reason, we run long
simulations for over 300 periods ($\sim$ 1000 time units) after disregarding the initial transient time
and record every integration step for higher accuracy.
Four sets of simulations were analyzed for each healthy and patient subjects using the FHN 
and the LIF models. In all cases, detailed recording over time were gathered from representative voxels. 
\par In the following subsections
typical voxels will be depicted in Figs. \ref{fig:10} and \ref{fig:11}. 
For the healthy subject H(1), Fig.~\ref{fig:01}a, the presented voxels (i,j,k) are: (48,93,19) 
front left; (77,136,19) inside the
cc area; (132,136,19) centrally located in the figure; (138,156,19) inside the cc area and (179,133,19) 
back front left area. For the patient P(1), Fig.~\ref{fig:01}b,
the presented voxels are: (36,85,19) 
front left; (61,104,19) inside the
tumorous area; (80,131,19) inside the tumorous area; (133,150,19) inside the cc area and (169,91,19) 
inside the cc area. Note that different voxels are used in each case, because the cc
areas in the diseased subject were delocalized due to the presence of the tumor.
\par The time series recordings were Fourier transformed to detect the dominant 
frequencies and the results on subjects H(1) and P(1) of the
FHN and LIF simulations are
presented in Sec.~\ref{sec:spectra-FHN} and Sec.~\ref{sec:spectra-LIF}, respectively. 
Additional results on the spectra of
 subjects H(2) and P(2) are included in the Supplementary Material (Multimedia view).
\subsection{FHN Spectra}
\label{sec:spectra-FHN}
\par
\begin{figure}[h]
\includegraphics[width=0.55\textwidth,angle=0.0]{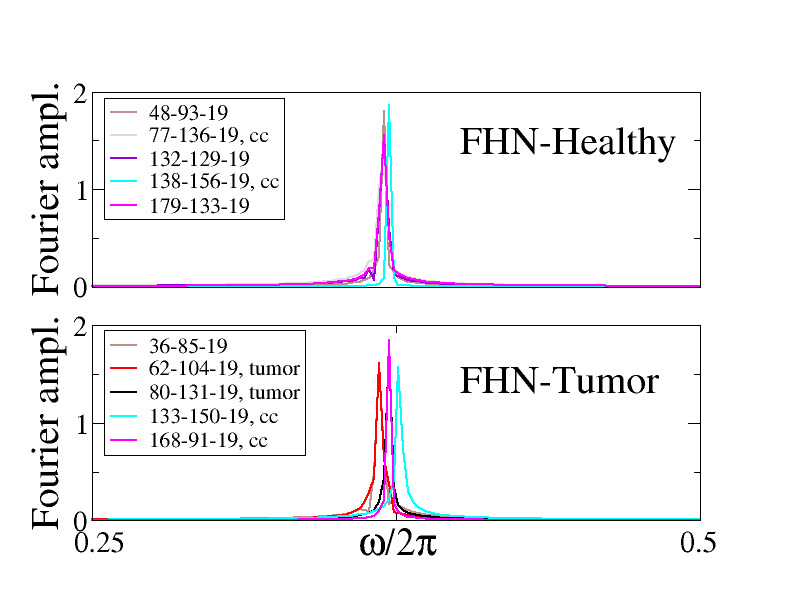}
\caption{\label{fig:10} 
  Fourier spectra of various voxels from FHN simulations. Top: Healthy subject H(1). Bottom: Patient P(1). 
The various colors correspond to the different voxels. In the subject with tumor the
black and red lines correspond to voxels from the tumor area. The voxels denoted by cc are located
in the corpus callosum areas. The spectra are computed over 300 cycles of the FHN dynamics.
Other parameters as in Fig.~\ref{fig:02}.
}
\end{figure}

 The Fourier spectra of the FHN dynamics in the neuron axons network of the healthy subject H(1)
 are presented in
the top panel of Fig.~\ref{fig:10}. All nodes (voxels), independently of position, demonstrate the same frequency
characteristics. Even the nodes residing in the cc areas demonstrate the same characteristics of the
basic frequency component as the other
nodes in the white matter, while small deviations are noted in the higher harmonics (not shown).
\par A distinct difference is noted when there is a tumorous area as depicted in the lower panel
of Fig.~\ref{fig:10}, which presents the Fourier spectra of the individual nodes
 when the connectivity matrix of the patient P(1) is used. In this case,
the timeseries originating from different areas develop different collective frequencies, although 
in the simulations all elementary oscillators have identical natural frequencies. 
This difference in the spectra can be attributed to the 
inhomogeneity that the tumor induces in the brain and confirms the observation of the chimera-like states 
which was reported in the previous sections \ref{results:anomalous-FHN} and \ref{results:anomalous-LIF}.
The different contributions visible on the basic frequency (Fig.~\ref{fig:10}, lower panel)
 are accentuated on the first harmonic (not shown).

\par For the simulations in the present subsection, all parameters used are the same as in Fig.~\ref{fig:02}.
Further studies in this direction, including the use of different FHN parameters,
may enlarge the frequency difference in the tumorous brain and help to identify the
affected regions even when their size is small.

\subsection{LIF Spectra}
\label{sec:spectra-LIF}
\par
\begin{figure}[h]
\includegraphics[width=0.55\textwidth,angle=0.0]{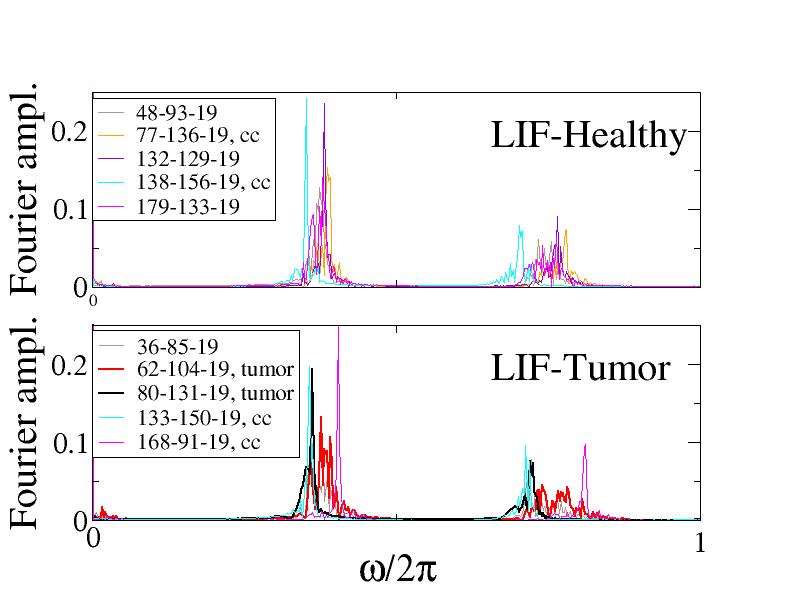}
\caption{\label{fig:11} 
  Fourier spectra of various voxels from LIF simulations. Top: Healthy subject H(1). Bottom: Patient P(1). 
Line colors and other notation as in Fig.~\ref{fig:10}.
The spectra are computed over 300 cycles of the LIF dynamics. All parameters as in Fig.~\ref{fig:04}.
}
\end{figure}

 The Fourier spectra when the LIF dynamics are used for the integration of the potential in the healthy
and tumorous case are depicted in Fig.~\ref{fig:11}. The top (bottom) panel depicts the Fourier spectra of the healthy (patient)
subject.
\par In both cases the spectra do not present a single frequency but a continuous band of frequencies,
both around the basic frequency as well as its harmonics. For this reason it is not possible to clearly
identify differences in the Fourier spectra between the voxels belonging to the tumorous regions and
other non-tumorous areas. The results do not improve when longer time series (as long as $\sim$ 19000 time 
units) are considered. 
\par These results give a first evidence that the FHN model dynamics may be more
appropriate than the LIF model for distinguishing between healthy and tumorous
areas in the brain when the Fourier spectra are used. Nevertheless, by adjusting 
parameters, such as $\alpha$ or $\phi$ in the FHN model, $u_{\rm th}$ in the LIF model and $h, \> R$
in the connectivity matrix we may achieve better discrimination levels.

\section{Conclusions}
\label{sec:conclusions}

\par Simulations inspired by the connectivity of the neuron axons network
were conducted, based on MRI-DTI data from
healthy and tumor suffering brains. Two neuronal 
models were used to simulate the dynamics in the individual
voxels recorded from the MRI images, the FitzHugh-Nagumo and the Leaky Integrate-and-Fire models.  
In the healthy brain
both models consistently differentiate between the corpus callosum regions, which are regions with high
density of neuron axons, and the rest of the white matter. In tumorous brains, the destructed regions are
clearly visible in the mean phase velocity diagrams of both models. 

\par The numerical results demonstrate that the healthy brain presents chimera-like states where regions
with high white matter concentrations in the direction connecting the two hemispheres
 act as the coherent domain, while the rest
of the brain exhibits incoherent oscillations. To the contrary, in brains with destructed 
white matter structure 
traveling waves are produced initiated around the region where the tumor is located. These areas act
as the pacemaker of the waves sweeping across the brain.

\par The present simulations, using specific parameter variables, give a first evidence of the pacemaker effect
and the chimera-like states. It would be useful to scan further the parameter space of the FHN and LIF
models to determine the parameter regions where these effects are accentuated. 
Further studies with more subjects and different sizes of tumors need to be assessed, to verify
whether the pacemaker effect could be useful as indicator of brain tumorous 
states, with particular reference in cases of small, difficult to detect  abnormalities.

\par 
In this study we have tested the FHN and LIF dynamics and have
shown that both models give 
consistent results.  In future studies other neuronal models 
can be implemented to evaluate the universality of these results across models.
 In particular, neuronal population dynamics models can be employed to represent the local dynamics
on a single voxel and test whether this class of models leads to comparable conclusions.

\par In a different network approach, connectivity matrices might be constructed using
nonlocal interactions over variable ranges. The properties of these matrices can be compared 
between control subjects and patients and these could also be examined as indicators/biomarkers of  tumor. The form and statistics of the constructed (nonlocal) networks may also give some
insights in the synchronization properties observed in the patients and the control subjects.

\par Apart for the cases of tumors, similar numerical studies might be conducted using neuron axons
networks from MRI studies of brains suffering from Alzheimer, Parkinson or other disorders
which affect the structure of the brain neurons and their connectivity networks.

\section*{Supplementary Material}
See in the  supplementary material  the eight videos concerning the four subjects as follows
a) FHN simulations of the healthy control brains (subjects H(1) and H(2)),
b) FHN simulations of the patient (tumor) brains (subjects P(1) and P(2)),
c) LIF simulations of the healthy control brains (subjects H(1) and H(2)), 
d) LIF simulations of the patient brains (subjects P(1) and P(2)).

Additional images are also uploaded as
e) Fourier spectra of subjects H(2) and P(2) with FHN dynamics and
f) Fourier spectra of subjects H(2) and P(2) with LIF dynamics.

Collections of 15 consecutive profiles are provided for each of the following cases:
g) FHN simulations of subject H(1), 
h) FHN simulations of subject P(1),
i) LIF simulations of subject H(1) and
j) LIF simulations of subject P(1).

\begin{acknowledgments}
I.K. and A.P. acknowledge helpful discussions with Dr. J. Hizanidis.
D.A.V. gratefully acknowledges logistic and financial support by the 
Association of Friends of Children with Cancer ``ELPIDA''.
I.O., A.Z and E.S. acknowledge financial support by the Deutsche Forschungsgemeinschaft 
(DFG, German
Research Foundation) - Projektnummer - 163436311 - SFB 910. A.P. acknowledges support 
of this work via the project MIS 5002567, implemented under the “Action for the Strategic 
Development on the Research and Technological Sector”, funded by the Operational 
Programme "Competitiveness, Entrepreneurship and Innovation" (NSRF 2014-2020) 
and co-financed by Greece and the European Union (European Regional Development Fund).  
This work was supported by computational time granted from the 
Greek Research \& Technology Network (GRNET) in the National HPC facility - ARIS, 
under project CoBrain4 (project ID: PR007011).

\end{acknowledgments}

\section*{Data Availability Statement}
The data that support the findings of this study are available from the corresponding author upon reasonable request.

\medskip\medskip
\bibliography{./provata.bib}
\end{document}